\documentclass[11pt,aps,prd,preprint,nofootinbib,superscriptaddress]{revtex4-1}
\usepackage{amsmath}
\usepackage{epsf}
\usepackage{graphicx}  
\usepackage{epstopdf}
\usepackage{dcolumn}   
\usepackage{bm}        
\usepackage{color}
\usepackage[english]{babel}
\usepackage{wasysym}
\begin{document}
%
\title{Constraints on the radiation temperature before inflation}
\author{Ram\'{o}n Herrera\footnote{E-mail: ramon.herrera@pucv.cl}}
\affiliation{Instituto de F\'{\i}sica. Pontificia Universidad Cat\'{o}lica de Valpar\'{\i}so,
casilla 4056, Valpara\'{\i}so, Chile}
\author{Diego Pav\'{o}n\footnote{E-mail: diego.pavon@uab.es}}
\affiliation{Departmento de F\'{\i}sica, Universidad Aut\'{o}noma de Barcelona,
08193 Bellaterra (Barcelona), Spain}
\author{Joel Saavedra\footnote{E-mail: joel.saavedra@pucv.cl}}
\affiliation{Pontificia Universidad Cat\'{o}lica de Valpar\'{\i}so, Chile}
\begin{abstract}
\noindent We consider the short period of cosmic expansion ranging from the end of the Planck era to the beginning
of inflation and set upper and lower limits on the temperature of the radiation at the commencement
of the inflationary phase.
\end{abstract}
\maketitle
\section{Introduction}\label{introduction}
\noindent Nowadays it is widely accepted that the homogeneity and isotropy   we observe today at cosmic scales, as well as
the seeds of the present matter structure, can be explained by a short period of very rapid expansion experienced
by the primeval universe. According to the common lore this expansion was driven by a scalar field (not yet identified),
whose nearly constant potential dominated every other form of energy during the aforesaid expansion (see \cite{Basset2006} for a
comprehensive review), though other mechanisms responsible for inflation are conceivable \cite{1980aleksey,2011salvatore}. Either way,  this
expansion erased, or nearly erased, any information we could otherwise acquire about the state of the universe
before inflation set in. Nevertheless it is obvious that the main energy component in the interval between the
end of the Planck era and the commencement of the inflationary period fulfilled all the energy conditions and it seems natural to
identify it as radiation (photons and neutrinos) plus ultrarelativistic matter, all of them at thermal equilibrium with
each other at temperature $T_{\gamma}$. The latter goes down with expansion as the inverse
of the scale factor of the Friedmann-Robertson-Walker (FRW) metric. Ihis paper aims to set limits
(lower and upper bounds) on that temperature at the onset of inflation.
\\  \

\noindent To do this we shall assume that: (i) The inflationary expansion began below the Planck energy scale.
(ii) Inflation was driven by some scalar field, $\phi$, that violated the energy conditions, had vanishing 
entropy as it was in a pure quantum state. (iii) During that short inflationary period the scalar potential, 
$V(\phi)$, was the dominant energy component. (iV) Between the Planck regime and the beginning of inflation, 
radiation plus ultrarelativistic particles dominated the expansion.
\\  \

\noindent Clearly, the said temperature limits  are to depend on the inflaton field. We shall consider two well motivated cold inflationary
scenarios consistent with the 2015 Planck's data \cite{2015planck}. First  the chaotic model \cite{linde-plb-1983}; secondly,
Chiba's model \cite{2015chiba}. The latter covers several interesting models as limiting cases and it is rater general.
\\  \

\noindent In both models we shall resort to the generalized second law of thermodynamics. This law was introduced in
connection with the thermodynamics of black holes \cite{bekenstein} and it is a straightforward extension to gravitational physics
of  the ordinary second law when the black hole entropy is considered. It lays that  the entropy of a black hole
(one quart of the area of its event horizon) plus the entropy of its surroundings cannot diminish. In cosmic settings,
and in the absence of black holes, it has been formulated by saying that the entropy of the cosmic horizon (either the
future event horizon or the apparent horizon)  plus the entropy of matter and fields inside the horizon cannot
decrease \cite{gibbons}. Recently, it has been applied in the study of the late evolution of some FRW models 
\cite{2013zp-diego}, and in the analysis of the viability of nonsingular bouncing universe
models \cite{2016ferreira}.
\\   \

\noindent Here we shall consider the apparent horizon, defined as the boundary hypersurface of the antitrapped spacetime
region, since it always exists both for expanding and contracting universes. By contrast, the future event horizon
exists only if the universe accelerates for  $\, t \rightarrow \infty$. A spacetime region is called antitrapped if the
ingoing and outgoing radial null geodesics have positive expansion (see \cite{bak-rey} and  \cite{cai-cao} for details).
As a consequence the radius of the apparent horizon (centered at the comoving observer position) reads
$r_{H} \equiv (H^{2} \, + k \, a^{-2})^{-1/2} \, $, where $k$ denotes the scalar curvature index of the FRW metric and, as usual,
$H = \dot{a}/a$, the Hubble factor. The entropy of the horizon, proportional one quarter of the area of the horizon,
is simply $\, S_{H} = \pi \, r^{2}_{H}/l^{2}_{{\rm p}}$ in units of the Boltzmann constant, with $l_{{\rm p}}$ the Planck
length.  At the primeval epochs we are interested in (before inflation and just after it) $\, a  \ll 1 \,
$ while $\, H \, $ is large, thereby the horizon entropy is excellently approximated by
$\, S_{H} = \pi \, l^{-2}_{{\rm p}}\, H^{-2}$. In this work we shall use units such that $k_{B} = c = \hbar = 1$.

\section{Upper and lower bounds on the temperature at the beginning of cold inflation}
\noindent Let us consider that inflation started somewhat below the Planck energy scale. Then, it is reasonable to assume
that before the inflaton field dominated the expansion this was driven  by some other
energy fields that comply with the  null, dominant and strong energy conditions. Let us model these fields
as thermal radiation characterized by some common temperature $T_{\gamma} \propto a^{-1}$.
\\  \

\begin{figure}[htbp]
    \centering
        \includegraphics[width=0.8\linewidth]{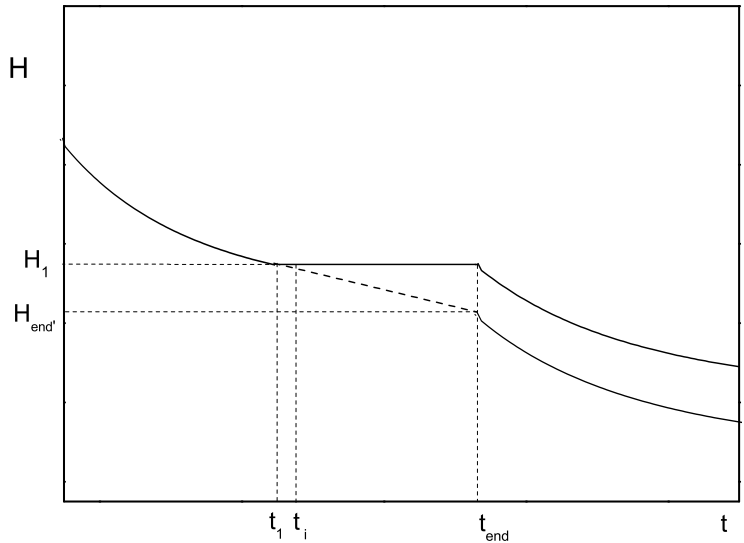}
\caption{{\small Schematic evolution of the Hubble factor with time. Inflation starts at $t_{1}$  and ends at $t_{{\rm end}}$.
Before $t_{1}$ the universe was dominated by fields that fulfill the energy conditions whence $\, \dot{H} < 0$. If the inflationary
expansion were exactly de Sitter (solid horizontal line), then $H_{1} = H_{{\rm end}}$. The long-dashed line  (connecting
$(t_{1}, H_{1})$ and $(t_{{\rm end}}, H_{{\rm end'}}) $) describes more realistically the inflationary stage rendering it
now compatible with the second law (see the main text). In this scenario, the entropy of the apparent horizon augments during this epoch. The instant
$t_{i}$ corresponds to the time when the energy density of radiation has declined due to the inflationary expansion by a factor
of, say, 100 \textemdash about 5 e-folds. From $\, t_{{\rm end}}$ onwards the universe gets dominated by the matter and
radiation generated nearly instantaneously in the reheating, again $H \propto t^{-1}$.}}
\label{fig:infl1}
\end{figure}

\noindent Let the de Sitter inflation ($H =$ constant) begin at $t = t_{1}$ and end at $t = t_{{\rm end }}$, \textemdash see Fig. \ref{fig:infl1}.
The radiation entropy inside the apparent horizon at $t_{1}$ will be $\, S_{\gamma}(t_{1}) = K_{2} \, T^{3}_{\gamma 1}/H^{3}_{1}$,
where $K_{2} = (4 \pi/3) (2 \pi^{2} g_{*}/45)$ with $\, g_{*}$ the number of relativistic degrees of freedom, approximately $\, 100$
in our case. Because of the temperature redshift,  $T_{\gamma \, {\rm end}} = T_{\gamma 1}/\exp(H (t_{{\rm end}} - t_{1})) \ll T_{\gamma 1}$,
the corresponding entropy at $t_{{\rm end}}$ will be  negligible as compared to  $S_{\gamma}(t_{1})$. On the other hand, since
$H_{1} = H_{{\rm end}}$, in this inflationary scenario the total entropy (i.e., the entropy of the apparent horizon,
$\propto H^{-2}$, plus the entropy of the radiation inside the horizon) will have diminished at $t_{{\rm end}}$ by about
$S_{\gamma}(t_{1})$. Bearing in mind the generalized second law, this means that the expansion rate, $\, H$, cannot be
exactly constant during inflation. It must experience a slight decrease so that $ \, H(t_{{\rm end}})\, $ should be
in reality  $\, H_{{\rm end'}} = H_{1} (1\, -\, x)$ with $\, 0 < x \ll 1$, a small number fixed
by the values of $\, H_{1}  \,$ and $ \, H_{{\rm end'}}$.
\\  \

\noindent From $\, \Delta S_{H} \geq \Delta S_{\gamma}\, $ it follows that
\begin{equation}
\frac{\pi}{l^{2}_{{\rm p}}}\, \left(\frac{1}{H^{2}_{{\rm end'}}} \, - \, \frac{1}{H^{2}_{1}}\right)
\geq K_{2} \frac{T^{3}_{\gamma 1}}{H_{1}^{3}}.
\label{Delta1}
\end{equation}
\\   \

\noindent If the inflationary expansion is driven  by a scalar field, $\, \phi \, $, that slowly rolls-down its potential  $V(\phi)$,
with the help  of Friedmann's equation $\, H^{2} = \kappa\, V(\phi)$, Eq. (\ref{Delta1}) can be recast as
\begin{equation}
T_{\gamma 1}  \leq  \left\{2 \,  \frac{K_{1}}{K_{2}}\, \sqrt{\kappa} \, [V^{1/2}(\phi_{i}) \, - V^{1/2}(\phi_{{\rm end}})] \, m^{2}_{p} \right\}^{1/3},
\label{Delta2}
\end{equation}
where $\, \kappa = 8 \pi/(3 m^{2}_{p})$  (recall that in our units $\, G = l^{2}_{p} = 1/m^{2}_{p}$), $\, K_{1} = 2 \pi$, and $\, K_{2} = 183.46$
for $\, g_{*} = 100$.  We have substituted $\phi(t_{i})$ by $\phi(t_{1})$ where $t_{i} \gtrsim t_{1}$ is the time at which the
radiation has redshifted by a factor of 100, which corresponds to approximately 5 e-folds. At that time the energy density is
already overwhelmingly  dominated by the scalar field. Not so at $t_{1}$ when inflation began, because at that time
the energy density of the radiation \textemdash though lower than $V(\phi_{1})$ \textemdash was, however,
of the same order.
\\  \

\noindent To go from (\ref{Delta1}) to (\ref{Delta2}) we used $H_{1} \, x = H_{1} \, - \, H_{{\rm end'}}$  as well as
\begin{equation}
H^{3}_{1} \left(\frac{1}{H^{2}_{{\rm end}}}\, - \, \frac{1}{H^{2}_{1}}\right) \simeq 2 H_{1} \, x \, \qquad(0 < x \ll 1).
\label{2H1x}
\end{equation}
Note that $\; x = 1 \, - \, \sqrt{V(\phi_{{\rm end}})/V(\phi_{i})}$. Also, $\, H_{1} \gtrsim H_{i} =
\sqrt{\kappa \, \, V_{i}(\phi)}$, and $\, H_{{\rm end'}} =  \sqrt{\kappa \, \, V_{{\rm end}}(\phi)}$.
Next we apply Eq. (\ref{Delta2}) first to the chaotic model of inflation and then to Chiba's model.

\subsection{Chaotic inflation}
\noindent  The simplest chaotic inflation model \cite{linde-plb-1983} is characterized by the potential
\begin{equation}
V(\phi) = \frac{1}{2}\, m^{2} \, \phi^{2},
\label{chaoticpot}
\end{equation}
where $\, m \, $ stands for the inflaton mass. To set an upper bound on $T_{\gamma}$ at the beginning
of inflation it is expedient to express this potential in terms of the number of e-folds. The latter
is given by
\begin{equation}
N = \int_{t_{1}}^{t_{{\rm end}}}{H \, dt} = 3 \, \kappa \, \int_{\phi_{{\rm end}}}^{\phi_{i}}{\frac{V}{V_{, \phi}}\, d \phi}.
\label{Nefolds}
\end{equation}
Therefore,
\[
\frac{m^{2}}{2}\, N = \frac{3}{4} \, [H^{2}_{i} \, - \, H^{2}_{{\rm end'}}] \, , \quad \phi_{{\rm end}} = \frac{2}{\sqrt{6 \, \kappa}}\, ,
\]
and
\[
H_{i} = \sqrt{\frac{2}{3}\, m^{2} \left(N \, + \, \frac{1}{2}\right)} \lesssim H_{1} \, ,
\]
\noindent Consequently, assuming $N = 60\, $ we have from Eq. (\ref{Delta2}) with $K_{1 }/K_{2} \simeq 0.0343$
\begin{equation}
T_{\gamma 1} \leq \left\{2 \times  0.034 \, \left[\sqrt{\frac{2}{3}\left(60 + \frac{1}{2}\right)} \,
- \, \frac{1}{\sqrt{3}}\right] \, m \, m^{2}_{{\rm p}}\right\}^{1/3}.
\label{Delta3}
\end{equation}
\\   \

\noindent On the other hand, Planck's experiment yields  $\, P_{s} \sim 10^{-9}$  for the power of scalar
modes at the end of inflation. This for the present model implies  $\, m \simeq 10^{-6} \, m_{{\rm p}}$
\cite{Basset2006}.  Plugging this into (\ref{Delta3}) leads to an  upper bound on the radiation
temperature at the start of inflation, namely, $T_{\gamma 1} \leq 7.3 \times 10^{-3} \, m_{{\rm p}}$.
\\  \

\noindent  A lower bound on $T_{\gamma 1}$ can be obtained as follows. Inflation began when $\ddot{a}$
became positive, i.e., when $\rho_{\gamma} + \rho_{\phi} + 3 (p_{\gamma} + p_{\phi})$
vanished. Recalling that at $t = t_{1}$ one had that $(\dot{\phi})^{2}/2$ was negligible against $V(\phi)$ and $\rho_{\phi}$,
because of the slow roll, and that $p_{\gamma} = \rho_{\gamma}/3$, $\rho_{\phi} \simeq V(\phi)$, $p_{\phi} \simeq - V(\phi)$,  this
occurred when $\rho_{\gamma} = V(\phi)$, i.e.,  $\rho_{\gamma 1} = V(\phi_{1})$. On the other hand, $V(\phi_{1}) > V(\phi_{{\rm end}})$.
Therefore $\rho_{\gamma 1} > V(\phi_{{\rm end}})$; that is to say,
\begin{equation}
\frac{\pi^{2}}{30} \, g_{*}\, T^{4}_{\gamma 1} > \frac{1}{2} \, m^{2} \, \phi^{2}_{{\rm end}} =  \frac{1}{3} \, \frac{m^{2}}{\kappa}\, .
\label{lowerbound1}
\end{equation}
Keeping in mind that  $\, m \simeq 10^{-6} \, m_{{\rm p}}$ and that $g_{*} = 100$, we get the lower bound   \newline
$\, T_{\gamma 1} > 1.8 \times 10^{-4} \, m_{{\rm p}}$.\\  \
\noindent Combining this with the upper bound derived above, we can write \newline
$1.8 \times 10^{-4} \, m_{{\rm p}} < T_{\gamma 1} < 7.3 \times 10^{-3} \, m_{{\rm p}}$.

\subsection{Chiba's model}
\noindent In this model  the potential expressed in terms of the number of e-folds reads \cite{2015chiba}
\begin{equation}
V(N) = \frac{N}{\alpha \, + \, \beta N},
\label{VN}
\end{equation}
where $\, \alpha\, $ and $\, \beta$ are integration constants with $\, \alpha > 0$ and  $\, \beta < \alpha $ except that
$\beta \neq 0$. Both constants share units of $\, m^{-4}_{{\rm p}}$. The meaning of these becomes clear when one realizes
that $\, \alpha = - N^{2}\, d(V^{-1})/dN \,$ and $\, \beta = \lim_{N \rightarrow \infty} \, V^{-1}$. Note that,
at variance with most models of slow-roll, $V \, $ increases with $\, N$ (i.e., $d V(N)/d N = \alpha/(\alpha + \beta N)^{2} > 0$).
Next we briefly consider the cases $\, \beta >0 \, $ and $\, \beta  <0$. The case $\beta = 0$ reduces to the chaotic model
above considered \textemdash see \cite{2015chiba} for details.
\\  \

\noindent For $\beta > 0$ the number of e-folds can be written in terms of the inflaton field as
\begin{equation}
N(\phi) =  \frac{1}{\gamma^{2}} \, \sinh^{2}\, \left[\frac{\sqrt{3 \,\kappa}\, \gamma}{2}\, (\phi \, - \, C) \right] ,
\label{Nfi}
\end{equation}
where the parameter $\, \gamma \, $ is defined by $\, \gamma \equiv \sqrt{\mid \beta \mid /\alpha}$ and $\, C$
is an integration constant.
With  the help of (\ref{VN}) last equation produces
\begin{equation}
V(\phi) = \frac{1}{\beta} \tanh^{2}[\sqrt{3 \kappa}\, \gamma \, (\phi \, - \, C)/2],
\label{Vfi}
\end{equation}
This is known as the ``T-model" potential of Kallosh and Linde \cite{2013kallosh}. Likewise, when
$\, \gamma = \sqrt{2/3}$ it reduces to Starobinsky's model  \cite{1980aleksey}, and to the $\alpha$-attractor
model \cite{2013kallosh} for $ \, \alpha = 2/(3\, \gamma^{2}) $.
\\   \

\noindent For $\beta < 0 \, $ one has $ N = \gamma^{-2} \, \sin^{2}[(\sqrt{3 \, \kappa}/2)\, (\phi \, - \, C)]$, hence the condition
$\gamma^{2} \ll 1$ must be fulfilled for an efficient inflation. This combined with Eq. (\ref{VN}) yields
$V(\phi) \simeq (4 \alpha)^{-1} 3  \kappa\, (\phi \, - \, C)^{2}$, i.e., the quadratic potential.
\\   \

\noindent Since, contrary to the chaotic-inflation case, $d N/d \mid \phi \mid >0$,  when using Eq. (\ref{Delta2})
the order of the terms within the square parenthesis must be reversed
\begin{equation}
T_{\gamma 1} \leq \left\{2  \frac{K_{1}}{K_{2}}\, \sqrt{\kappa} \, [V^{1/2}(N_{{\rm end}}) \, - V^{1/2}(N_{i})] m^{2}_{{\rm p}}\right\}^{1/3}.
\label{Deltainvert}
\end{equation}
Using last expression  alongside Eq. (\ref{VN}) and choosing   $N_{i} = 5\, $ and $\, N_{{\rm end}} = 60$ we can write
\begin{equation}
\alpha^{1/6} \, T_{\gamma 1} \leq \left\{0.0343 \, \sqrt{\frac{8 \pi}{3}}\, \left[\sqrt{\frac{60}{1 \, + \, \gamma^{2} \, 60}} \,
- \, \sqrt{\frac{5}{1 \, + \, \gamma^{2} \, 5}}\right]\right\}^{1/3}\, m^{1/3}_{p}.
\label{TTmodel1}
\end{equation}
\\  \

\noindent  To estimate  $\, \gamma^{2}\, $ we exploit the fact that in the T-model  the tensor to scalar ratio, $r  \equiv P_{T}/P_{S}$,
between the power of tensorial and scalar modes generated at the end of inflation is given by  \cite{2015chiba}
\begin{equation}
r= \left\{\begin{array}{lc}
 \, \frac{8}{N_{{\rm end}} \, + \, \gamma^{2} \, N^{2}_{{\rm end}}}  & (\beta > 0) \\
\frac{8}{N_{{\rm end}}}  & (\beta \leq 0).
\end{array}\right.
\label{ratioperturbations}
\end{equation}
This combined with the result from the Keck Array and BICEP collaborations \cite{2016keck} about the  said ratio, namely
$r_{0.05} < 0.07$ at 95\% confidence level, yields $\, \gamma^{2} < 1.50793 \times 10^{-2}$ for $\, N_{{\rm end}} = 60$. \\
Now, taking into account that $\, V(\phi_{{\rm end}}) \simeq 10^{-12} \, m^{4}_{{\rm p}}$ \cite{2015planck} and that
$V(N = 60) \simeq 10^{2}/\alpha \, $ it follows that $\, \alpha \sim 10^{14} m^{-4}_{{\rm p}}$.
\\  \

\noindent Inserting the above estimation of $\, \alpha \, $ and the upper bound on $\gamma^{2}$ in Eq. (\ref{TTmodel1}) we obtain
\newline $T_{\gamma 1} \leq 3 \times 10^{-3} \, m_{{\rm p}}$. As it can be checked, this order of magnitude of the
bound is rather insensible to the initial number of e-folds, $N_{i}$, provided it remains low
(i.e., no larger than $10$). Likewise as it can be verified, for the specific case of $\beta = 0$
this upper bound remains practically unaltered.
\\  \

\noindent To determine  the lower bound on $T_{\gamma1}$ we proceed as in the case of chaotic inflation  and
use the experimentally found value $V(\phi_{{\rm end}}) \simeq 10^{-12} \, m^{4}_{{\rm p}}$; i.e.,
\begin{equation}
\frac{\pi^{2}}{30} \, g_{*}\, T^{4}_{\gamma 1} > 10^{-12} \, m^{4}_{{\rm p}} \, .
\label{lowerbound2}
\end{equation}
This gives $\, T_{\gamma 1} > 2.3 \times 10^{-4} \, m_{{\rm p}}$. In summary, the T-model implies \newline
$\, 2.3 \times 10^{-4} \, m_{p} < T_{\gamma1} < 3.05 \times 10^{-3} \, m_{{\rm p}}$. This interval is similar
to the one obtained in the case of the chaotic model.
\\    \

\noindent Equations (\ref{VN}) and (\ref{ratioperturbations}.a) combined with (\ref{Deltainvert}) lead to
\begin{equation}
r \geq \frac{3 \, K^{2}_{2}}{32 \, \pi K^{2}_{1}} \,  \frac{8 \, \alpha}{N^{2}_{{\rm end´}}} \, T^{6}_{\gamma 1} \, .
\label{r-upper-bound}
\end{equation}
This expression provides us with an upper limit on the ratio $\, P_{T}/P_{S}$. In arriving to it we have neglected the
term $V^{1/2}(N_{i})$. The latter can be estimated as three times less than  $ V^{1/2}(N_{{\rm end}})$; obviously
he approximation is not very good but we are just seeking a rough estimate for an upper bound on the said ratio. Inserting
for $T_{\gamma 1}$ the upper bound found above and $N_{{\rm end}} = 60$ in (\ref{r-upper-bound}) we find $ r \geq 10^{-8}$.

\section{Discussion}
\noindent At the moment no empirical information is available about the short expansion phase between the end of the Planck era and the beginning
of inflation. Maybe some  will be gathered in the future once the the gravitational waves supposedly produced in that period are detected and
successfully analyzed. Hence our current knowledge of that fleeting but crucial early epoch of the universe history is necessarily slim.
This lends much interest to  its study provided it is based on sound physics rather than just speculation.
\\  \

\noindent In this paper, we resorted to the
generalized second law of thermodynamics to set an upper bound on the temperature of the radiation at the beginning of
the inflationary expansion. A lower bound was found by imposing that the energy density of the radiation at that instant, $t_{1}$,
should be greater than that of the inflaton at $\, t_{{\rm end}}$, see Fig. \ref{fig:infl1}. We did this for
the simplest chaotic model and Chiba's model which show compatibility with the constraints derived from the Planck's mission \cite{2015planck}.
In both instances we got similar bounds on  $\, T_{\gamma 1}$ (roughly in the range $(10^{-4}, 10^{-3} \, m_{{\rm p}}$), consistent
with the very reasonable assumption  of being no lower than the reheating temperature. According to Chiba the latter lies in
the interval $\, [10^{-8}, 10^{-4} m_{p}]$ \cite{2015chiba}.
\\  \

\noindent Likewise, we obtained an upper limit of the tensor to scalar ratio, Eq. (\ref{r-upper-bound}). Though the bound is rather loose
it, nonetheless,  ensures that gravitational waves are indeed produced during inflation.

\acknowledgements{\noindent One of us, DP, is indebted to the ``Instituto de F\'{\i}sica  de la Pontificia Universidad
Cat\'{o}lica de Chile" for warm hospitality. This work was supported by the Agency CONICYT of the Chilean Government under contract
MEC 80160077.}


\end{document}